\documentclass[aps,prd,twocolumn,floatfix,nofootinbib]{revtex4-2}

\usepackage{graphicx}
\usepackage{amsmath}
\usepackage{amssymb}
\usepackage{xcolor}
\usepackage[colorlinks,citecolor=blue,linkcolor=blue,urlcolor=blue,anchorcolor=blue]{hyperref}
\allowdisplaybreaks
\newcommand\dx{{\rm d}}

\newcommand\etal{{\it et~al.}}

\newcommand\apjl{Astrophys. J. Lett.}

\newcommand\lrr{Living Rev. Relativity}
\newcommand\mnras{Mon. Not. R. Astron. Soc.}

\newcommand\pdu{Phys. Dark Universe}
\newcommand\plb{Phys. Lett. B}

\begin{document}

\title{Cosmological consequences of a scalar field with oscillating equation of state. II.
  Oscillating scaling and chaotic accelerating solutions}
\author{S. X. Tian}
\email[]{tshuxun@whu.edu.cn}
\affiliation{Department of Astronomy, Beijing Normal University, 100875 Beijing, China}
\affiliation{School of Physics and Technology, Wuhan University, 430072 Wuhan, China}
\date{\today}
\begin{abstract}
  Multiacceleration scenario can be used to solve the cosmological coincidence problem. In this paper, after considering the early radiation era, we revisit the cosmological dynamics of the oscillating dark energy model proposed in [\href{https://doi.org/10.1103/PhysRevD.101.063531}{Phys. Rev. D {\bf 101}, 063531 (2020)}]. We find this model allows the Universe evolves as oscillating scaling solution (OSS) in the radiation era and as chaotic accelerating solution (CAS) in the matter era. Mathematically, the transition from OSS to CAS is a route of period-doubling bifurcation to chaos. Physically, there are two reasons convince us that this scenario can be a nice picture to describe the real Universe. One is the global cosmological parameter constraints are practicable if the Universe evolves as OSS in the radiation era. The other is the late-time Universe described by CAS can successfully explain the observed cosmic acceleration at low redshifts.
\end{abstract}
\maketitle

\section{Introduction}
The cosmological coincidence problem is one of the biggest mysteries confounding cosmologists today \cite{Carroll2001.LRR.4.1,Peebles2003.RMP.75.559,Bull2016.PDU.12.56}. In the $\Lambda$ cold dark matter ($\Lambda$CDM) model, the effective dark energy density $\rho_\Lambda$ keeps constant while the pressureless matter density $\rho_{\rm matter}$ is proportional to $a^{-3}$ with the expansion of the Universe. The coincidence problem states why $\rho_\Lambda$ is comparable to $\rho_{\rm matter}$ at the present epoch. Theoretical explanations for this problem mainly focus on dynamical dark energy. For example, the dark energy models with tracker property, which allows higher dark energy density in the early Universe, can be used to alleviate the coincidence problem \cite{Peebles1988.ApJL.325.L17,Ratra1988.PRD.37.3406,Steinhardt1999.PRD.59.123504,Zlatev1999.PRL.82.896,Zhao2006.PLB.640.69}. However, these models cannot completely solve the problem. The reason is that the evolution of the dark energy relative energy density in these models has no essential difference from that in the $\Lambda$CDM model (see Fig. 11 in Ref. \cite{Carroll2001.LRR.4.1} for an illustration). The situation changes in the oscillating dark energy model with multiacceleration scenario. Using this scenario to solve the coincidence problem was first proposed by Nojiri and Odintsov \cite{Nojiri2006.PLB.637.139,Nojiri2006.PLB.639.144}, and recently re-proposed by us independently \cite{Tian2020.PRD.101.063531} (hereafter Paper I).

The model proposed in Paper I is a quintessence model with the potential
\begin{equation}\label{eq:01}
  V(\phi)=V_0\exp\left[-\frac{\lambda_1+\lambda_2}{2}\phi
  -\frac{\alpha(\lambda_1-\lambda_2)}{2}\sin\frac{\phi}{\alpha}\right].
\end{equation}
Inspired by dark energy models with single \cite{Copeland1998.PRD.57.4686} and double \cite{Barreiro2000.PRD.61.127301} exponential potentials, Paper I guessed and proved that the above potential can lead to the desired multiacceleration scenario. Quantitative discussion gives the viable parameter space of the model as $\lambda_1+\lambda_2>4$, $0<\lambda_2<0.39$, $\alpha=\mathcal{O}(1)$ and $V_0=\mathcal{O}(l_{\rm P}^{-2})$, where $l_{\rm P}$ is the Planck length\footnote{In principle, the possibility of $\lambda_2=0$ has not been ruled out. We adopt $\lambda_2>0$ to enhance the robustness of the model.}. Only one Planck scale parameter and three dimensionless parameters of order unity appear in the model's action. This is a key difference between our model and the model proposed in Ref. \cite{Nojiri2006.PLB.637.139}, and possibly makes our model a very natural physical theory (see more discussions in Paper I). However, Paper I also pointed out that the solution of the model with viable parameters falls into chaos in the matter era, which brings technical difficulty to the global cosmological parameter constraints and may make the model impractical.

In this paper, new mathematical property of the model proposed in Paper I is discovered: Period-doubling bifurcation. Furthermore, we point out that this new property can be used to eliminate the technical difficulty we discussed before. Our results are presented in two parts: Time domain and frequency domain.

\section{Properties in time domain}\label{sec:02}
In this paper, we follow the exact notation used in Paper I and we do not repeat the definitions. The Universe is assumed to be flat and contains radiation, pressureless matter and a canonical scalar field with the potential described by Eq. (\ref{eq:01}). The cosmic evolution equations can be written as
\begin{subequations}\label{eq:02}
\begin{align}
  \frac{\dx x_1}{\dx N}&=-3x_1+\frac{\sqrt{6}}{2}\lambda x_2^2+\frac{3}{2}x_1L,\\
  \frac{\dx x_2}{\dx N}&=-\frac{\sqrt{6}}{2}\lambda x_1x_2+\frac{3}{2}x_2L,\\
  \frac{\dx\lambda}{\dx N}&=\nu x_1,\label{eq:02c}\\
  \frac{\dx\nu}{\dx N}&=\frac{3x_1}{\alpha^2}(\lambda_1+\lambda_2-2\lambda),\label{eq:02d}
\end{align}
\end{subequations}
where $L=(1-w_{\rm m})x_1^2+(1+w_{\rm m})(1-x_2^2)$ and $w_{\rm m}$ is the EoS of normal matters. One constraint equation is
\begin{equation}\label{eq:03}
  \nu(\lambda)=\nu_\pm(\lambda)=\pm\frac{\sqrt{6}}{\alpha}\sqrt{\lambda(\lambda_1+\lambda_2)-\lambda^2-\lambda_1\lambda_2}.
\end{equation}
To test the model's ability to explain the cosmic late-time acceleration, Paper I numerically solved the above dynamic system with $w_{\rm m}=0$. In this section, to test the model's self-consistency across the whole cosmic history, we adopt
\begin{equation}\label{eq:04}
  w_{\rm m}(N)=\frac{1/3}{1+e^{N-N_{\rm eq}}},
\end{equation}
which gives the total EoS of radiation and pressureless matter. Here $N_{\rm eq}$ corresponds to matter-radiation equality and thus $N=N_{\rm eq}+8.13$ corresponds to now\footnote{Current observation gives the matter-radiation equality redshift $z_{\rm eq}=3400$ \cite{Aghanim2018.arXiv.1807.06209} and $\ln(1+z_{\rm eq})=8.13$. The value of $z_{\rm eq}$ is weakly dependent on the dark energy model. However, such slight difference does not affect our following discussions.}. Note that Eq. (\ref{eq:02}) applies to both constant and time-varying $w_{\rm m}$. We mainly use $w_{\rm tot}$, $\Omega_\phi$, $w_\phi$ and its first derivative
\begin{equation}
  \frac{\dx w_\phi}{\dx N}=\frac{2x_1x_2^2(\sqrt{6}\lambda\Omega_\phi-6x_1)}{\Omega_\phi^2}.
\end{equation}
to characterize the cosmological evolution. A viable dark energy model should be able to give $\Omega_\phi\approx0.7$, $w_\phi\approx-1$ and $\dx w_\phi/\dx N\approx0$ at present.

\begin{figure}[t]
  \centering
  \includegraphics[width=0.99\linewidth]{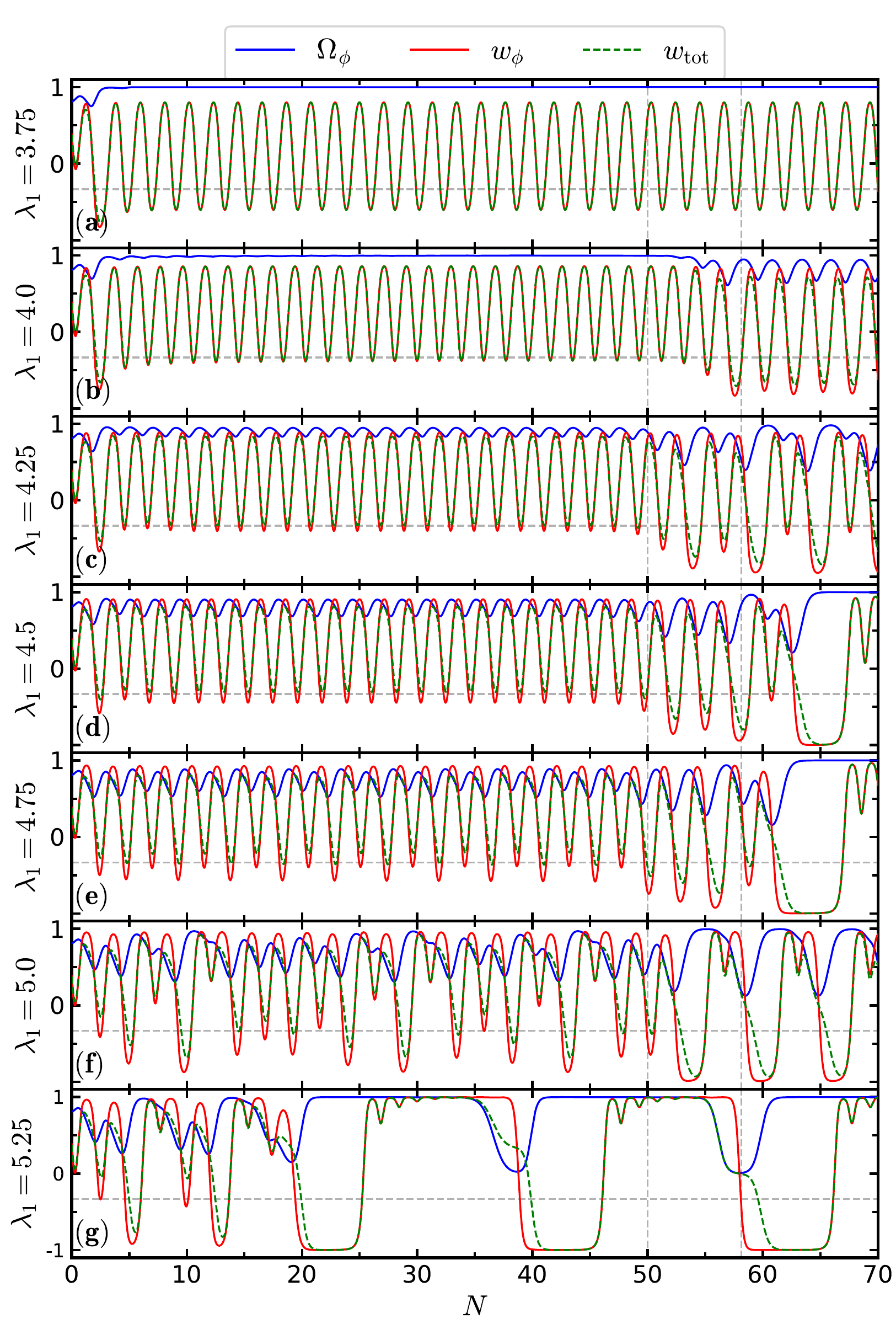}
  \caption{Evolution of the dark energy relative energy density $\Omega_\phi$, the EoS $w_\phi$ and $w_{\rm tot}$. The model parameters are $\lambda_1=3.75$, $4.0$, $4.25$, $4.5$, $4.75$, $5.0$, $5.25$, $\lambda_2=0.05$, $\alpha=0.6$ and $N_{\rm eq}=50$. The initial conditions are $x_{1,0}=0.75$, $x_{2,0}=0.5$, $\lambda_0=0.35$ and $\nu_0=\nu_+(\lambda_0)$. The first vertical dashed line corresponds to $N=N_{\rm eq}$ and the second one corresponds to $N=N_{\rm eq}+8.13$. The horizontal dashed line corresponds to $-1/3$. Note that $\ddot{a}/a=0$ if $w_{\rm tot}=-1/3$.}
  \label{fig:01}
\end{figure}

To get a first glance of the system's property, we plot the evolution of $\Omega_\phi$, $w_\phi$ and $w_{\rm tot}$ for the model parameter $\lambda_1=3.75$, $4.0$, $4.25$, $4.5$, $4.75$, $5.0$ and $5.25$ in Fig. \ref{fig:01}. Other parameter and initial condition settings can be found in the caption. Without loss of generality, all the numerical calculations in this paper start from $N=0$. When $\lambda_1=3.75$, i.e., $\lambda_1+\lambda_2<4$, the scalar field is always dominant over normal matters. Increasing $\lambda_1$ such that $\lambda_1+\lambda_2>4$ causes $\Omega_\phi$ to oscillate. This is consistent with the conclusion obtained in Paper I that $\rho_\phi$ is coincidence with $\rho_{\rm m}$ many times in the history requires $\lambda_1+\lambda_2>4$. When $\lambda_1\geqslant4.0$, we observe that the solution of Eq. (\ref{eq:02}) can be divided into two categories: Oscillating scaling solution (OSS) and chaotic accelerating solution (CAS). The OSS is periodic and can be regarded as a generalization of the classical scaling solution for the exponential potential \cite{Copeland1998.PRD.57.4686}. The CAS is chaotic. A key difference between these two solutions is that $w_\phi$ can be very close to $-1$ in CAS, but not in OSS. In CAS, both $w_\phi$ and $w_{\rm tot}$ oscillate between nearly $-1$ and $1$. When $\lambda_1\approx4.25$, the system evolves as OSS in the radiation era and as CAS in the matter era. When $\lambda_1\approx5.0$, the system evolves as CAS in both radiation and matter era. As we will see in the next section, the mathematical essence of the transition from OSS to CAS is a route of period-doubling bifurcation to chaos \cite{Strogatz2018.book}. For the OSS appeared in the radiation era, the minimum value of $w_{\rm tot}$ crosses $-1/3$ at $\lambda_1\approx4.75$. There are considerable separate accelerating phases in the radiation era if $\lambda_1\gtrsim4.75$. The influence of this property on the early Universe will be explored in the future.

We expect the realistic Universe to evolve as OSS in the radiation era and as CAS in the matter era. In this scenario, on the one hand, $w_\phi$ can be very close to $-1$ today and the model may be able to well fit the observations about the late-time Universe. On the other hand, if the Universe evolves as OSS in the radiation era, then we can use $N_{\rm eq}$ to replace the initial conditions of $\{x_1,x_2,\lambda,\nu\}$ in the global cosmological parameters constraints. This is because OSS is an attractor solution and different initial conditions only cause a phase difference of corresponding attractors. We can change the $N_{\rm eq}$ within one period with fixed initial conditions to reflect the influence of different initial conditions on the late-time cosmic evolution. This manipulation facilitates the cosmological parameter constraints.

\begin{figure*}[t]
  \centering
  \includegraphics[width=0.99\linewidth]{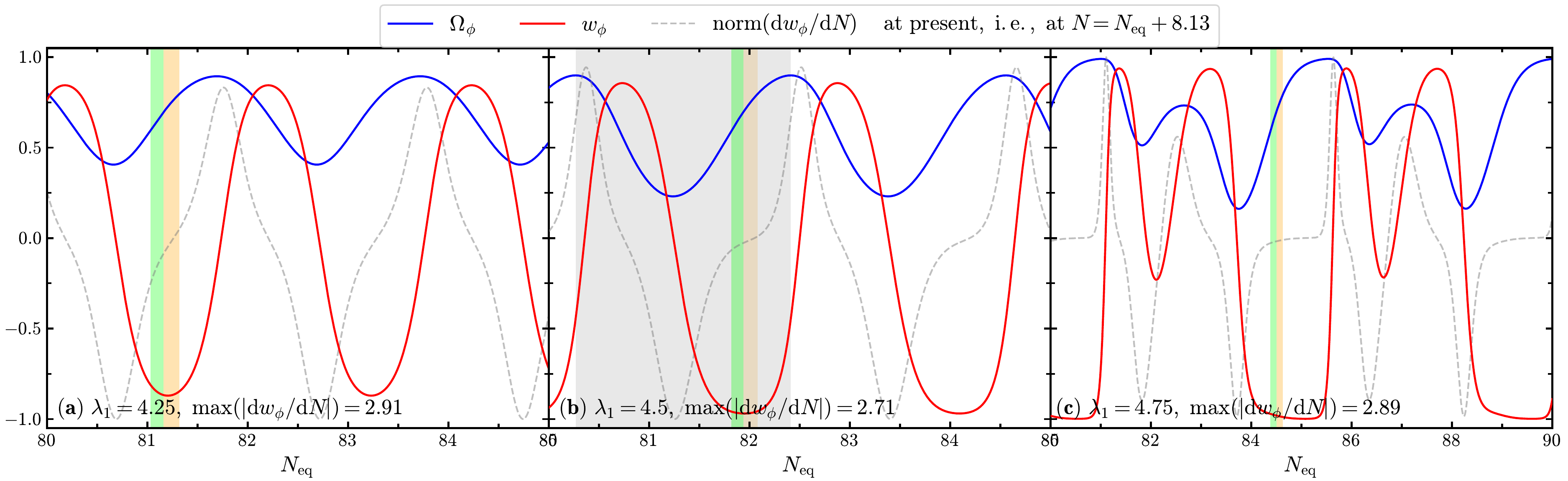}
  \caption{The values of $\Omega_\phi$, $w_\phi$ and $\dx w_\phi/\dx N$ at present versus the parameter $N_{\rm eq}$. For $\dx w_\phi/\dx N$, we plot its normalized value, which is defined as ${\rm norm}(\dx w_\phi/\dx N)\equiv(\dx w_\phi/\dx N)/{\rm max}(|\dx w_\phi/\dx N|)$. The model parameters are $\lambda_1=4.25$, $4.5$, $4.75$, $\lambda_2=0.05$ and $\alpha=0.6$. The initial conditions are all fixed at $x_{1,0}=0.75$, $x_{2,0}=0.5$, $\lambda_0=0.35$ and $\nu_0=\nu_+(\lambda_0)$. $0.6<\Omega_\phi<0.7$ in the green region and $0.7<\Omega_\phi<0.8$ in the orange region. In the subplot (b), the gray region corresponds to one period of the OSS attractor. There is a period-doubling bifurcation from $\lambda_1=4.5$ (b) to $\lambda_1=4.75$ (c) in the radiation era.}
  \label{fig:02}
\end{figure*}

Considering the strong dependence of the late-time cosmic evolution on the initial conditions, Paper I discussed the  technical difficulty of the cosmological parameter constraints. The result shown in Fig. \ref{fig:01} implies that the existence of OSS allows us to replace the initial conditions with $N_{\rm eq}$ in the parameter constraints. Here we test whether this replacement can solve the above technical difficulty. In Fig. \ref{fig:02}, we fix the initial conditions and plot the values of $\Omega_\phi$, $w_\phi$ and $\dx w_\phi/\dx N$ at present against $N_{\rm eq}$. These values are sufficient to characterize the evolution of the Universe at low redshifts. Parameter and initial condition settings can be found in the caption. For these parameter settings, Fig. \ref{fig:01} shows that the system evolves as OSS in the radiation era and as CAS in the matter era. Figure \ref{fig:02} shows that the values of $\Omega_\phi$, $w_\phi$ and $\dx w_\phi/\dx N$ at present change smoothly as $N_{\rm eq}$ changes. This means the posterior distribution also changes smoothly with respect to $N_{\rm eq}$ in the parameter constraints. Therefore the technical difficulty discussed in Paper I disappears if we replace the initial conditions with $N_{\rm eq}$ in the cosmological parameter constraints. In each subplot, the green region corresponds to $0.6<\Omega_\phi<0.7$ and the orange region corresponds to $0.7<\Omega_\phi<0.8$. In principle, there is no theoretical essential difference between $\Omega_\phi=0.6$ and $\Omega_\phi=0.8$ at present. We will not have more surprises if astronomical observations give $\Omega_\phi\approx0.6$ instead of $\Omega_\phi\approx0.7$ at present. The proportion of the colored region in one period in Fig. \ref{fig:02} is considerable. From this perspective, we do not need to fine-tune $N_{\rm eq}$ to explain the cosmic late-time acceleration.

\begin{figure}[t]
  \centering
  \includegraphics[width=0.99\linewidth]{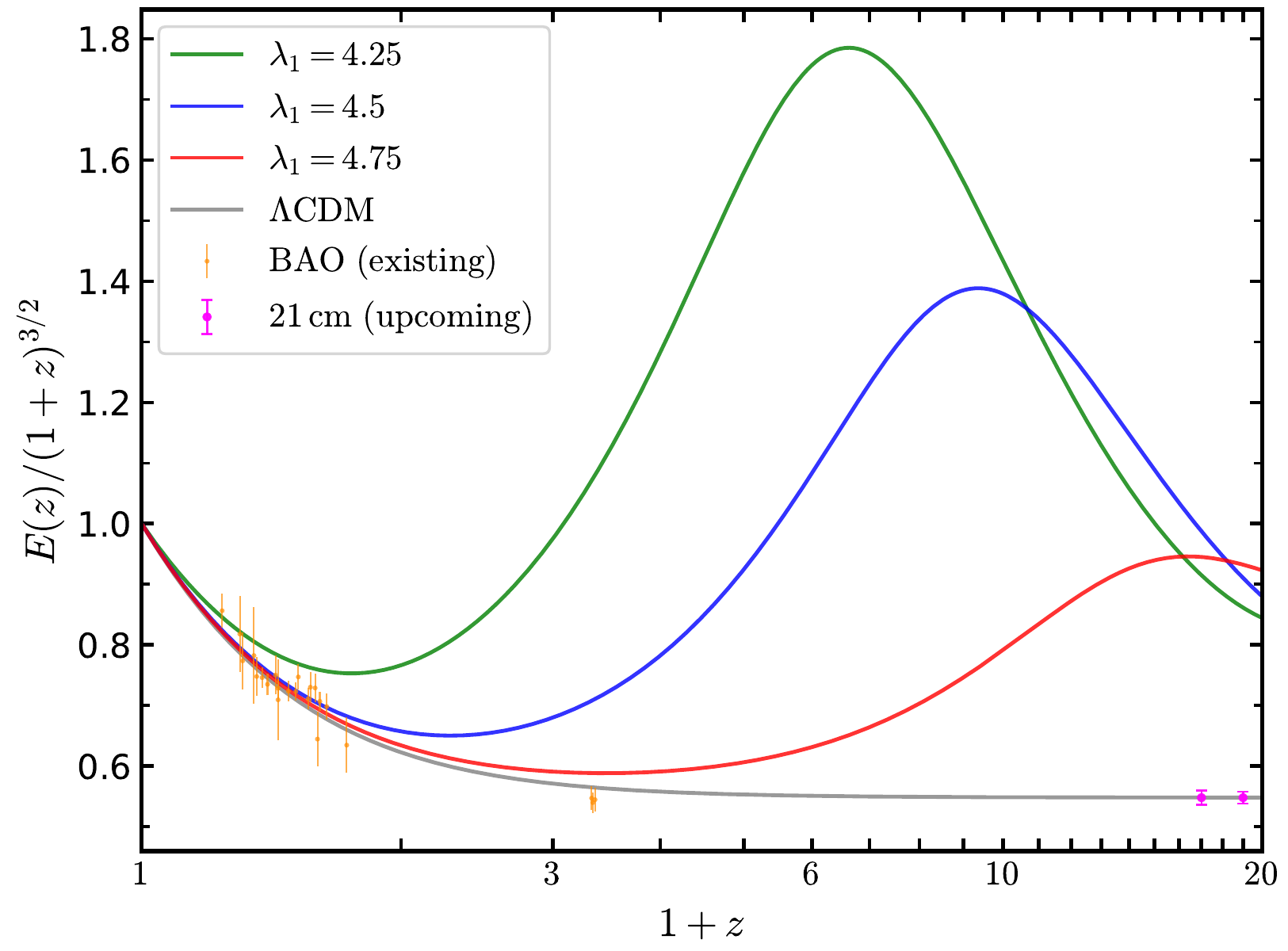}
  \caption{The Hubble parameter $H(z)$ as a function of redshift $z$. Note that $E(z)\equiv H(z)/H_0$. The model parameters are $\{\lambda_1,N_{\rm eq}\}=\{4.25,81.16\}$, $\{4.5,81.94\}$ and $\{4.75,84.50\}$. Other parameter and initial condition settings are the same as in Fig. \ref{fig:02}. The gray line corresponds to the standard $\Lambda$CDM model with $\Omega_\Lambda=0.7$. The orange points show the current measurements from BAO analysis (see Table IV in Ref. \cite{Li2019.arXiv.1911.12076} for a collection) and the red points show the projected measurements through 21\,cm observations \cite{Munoz2019.PRL.123.131301}. All the data points are normalized with the Planck result $H_0=67.4\,{\rm km}/{\rm s}/{\rm Mpc}$ \cite{Aghanim2018.arXiv.1807.06209}. Note that we do not need the value of $H_0$ when calculating the theoretical curves.}
  \label{fig:03}
\end{figure}

Figure \ref{fig:02} shows that $w_\phi$ is close to $-1$ and $\dx w_\phi/\dx N$ is close to $0$ when $\Omega_\phi\approx0.7$ in the colored region. This indicates that the model may be able to successfully explain the observed cosmic late-time acceleration. For a more rigorous testing, in Fig. \ref{fig:03}, we plot the Hubble parameter as a function of redshift for the model parameters corresponding to $\Omega_\phi=0.70$, $w_\phi\approx-1$ and $\dx w_\phi/\dx N\approx0$ in Fig. \ref{fig:02}. Detailed parameter settings can be found in the caption. The scatter points in Fig. \ref{fig:03} include current measurements of $H(z)$ through the baryon acoustic oscillations (BAOs; see Table IV in Ref. \cite{Li2019.arXiv.1911.12076} for a collection) and upcoming measurements through the 21\,cm signal from cosmic dawn \cite{Munoz2019.PRL.123.131301}. Figure \ref{fig:03} shows that the $\lambda_1=4.75$ case can fit the observations fairly well in the range of $z\lesssim2$, which covers most of the existing data points. We expect a better fit after slightly adjusting the parameters. Therefore, the model should be able to well fit the observations about the late-time Universe. The difference between our model and the standard $\Lambda$CDM model becomes significant at high redshifts. More importantly, this difference is observable through the 21\,cm signal. Future observations can be used to distinguish these two models. If the upcoming measurements of $H(z)$ at cosmic dawn are consistent with the $\Lambda$CDM model's predictions, then the multiacceleration scenario discussed in Paper I for solving the coincidence problem should be abandoned. However, it is also possible that 21\,cm observations bring more anomalies besides the global sky-average brightness temperature excess \cite{Bowman2018.Nature.555.67}.

\begin{figure}[t]
  \centering
  \includegraphics[width=0.99\linewidth]{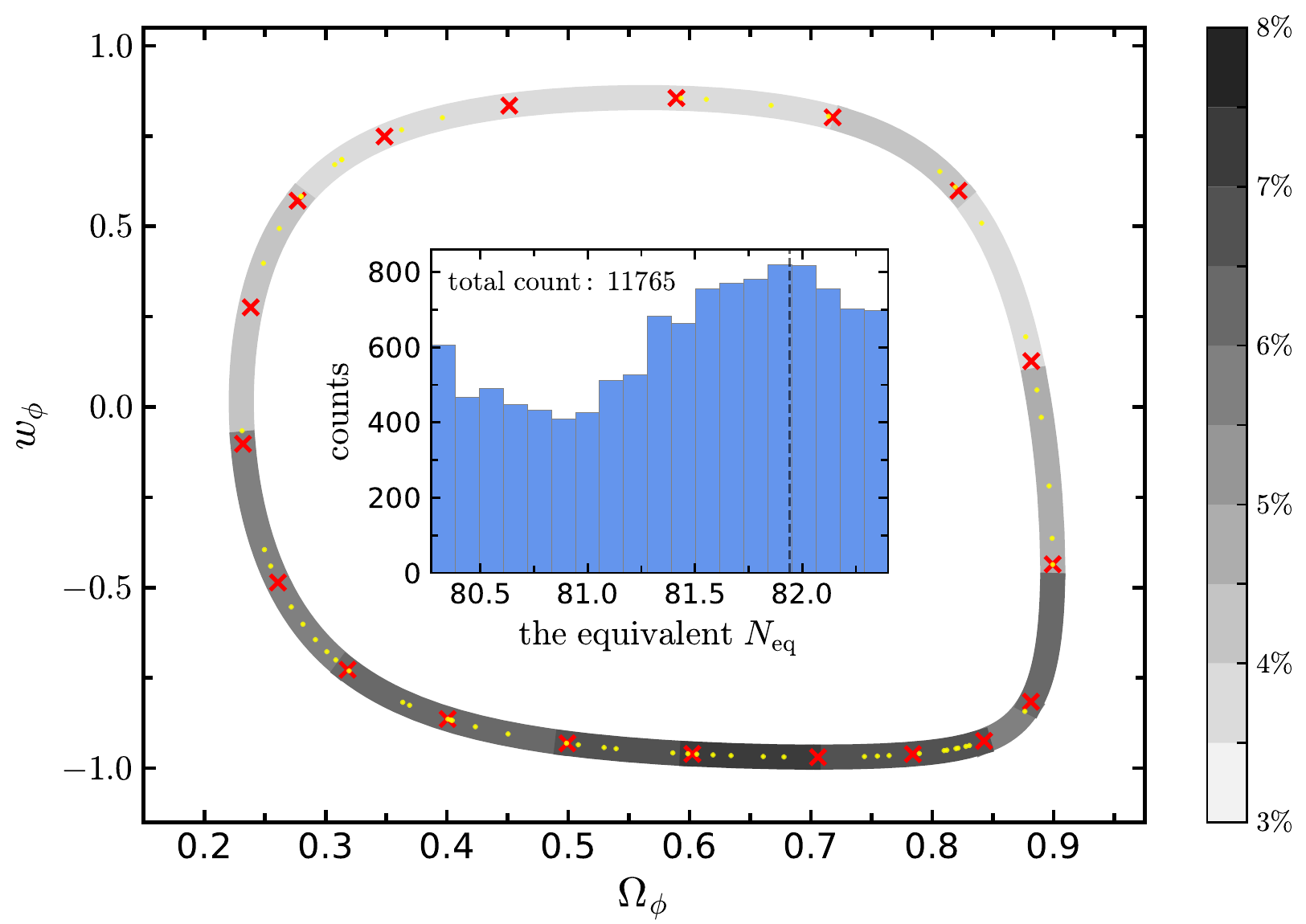}
  \caption{Theoretical prior of the equivalent $N_{\rm eq}$ (see the main text for model parameters and notations).}
  \label{fig:04}
\end{figure}

As we discussed before, replacing the initial conditions of $\{x_1,x_2,\lambda,\nu\}$ with $N_{\rm eq}$ facilitates the cosmological parameter constraints. For specific parameter settings about $\{\lambda_1,\lambda_2,\alpha\}$, the range of $N_{\rm eq}$ should be one period of the OSS attractor [see the gray region in Fig. \ref{fig:02} (b) for an illustration]. Furthermore, $N_{\rm eq}$ should have a theoretical prior distribution for the full range of the scalar field's initial conditions. Similar issues have been discussed in Refs. \cite{Marsh2014.PRD.90.105023,Garcia2020.PRD.101.063508} for other quintessence models. In Fig. \ref{fig:04}, we explore this theoretical prior for the model parameters studied in Fig. \ref{fig:02} (b). Figure \ref{fig:04} is plotted as follows:
\begin{enumerate}
  \item We divide the gray region of Fig. \ref{fig:02} (b) equally into 19 bins along the $x$-axis. We plot the $\Omega_\phi-w_\phi$ circle based on these 19 bins and the boundaries are indicated by red crosses. The $x$-axis of the gray region in Fig. \ref{fig:02} (b) is referrd as the equivalent $N_{\rm eq}$ in the inserted histogram.
  \item We fix the model parameters as $\lambda_1=4.5$, $\lambda_2=0.05$, $\alpha=0.6$ and $N_{\rm eq}=80$. The initial conditions are randomly given by the uniform distribution and $x_{1,0}\in[-1,1]$, $x_{2,0}\in[0,1]$ with $x_{1,0}^2+x_{2,0}^2<1$, $\lambda_0\in[\lambda_2,\lambda_1]$, $\nu_0=\nu_+(\lambda_0)$ with the probability of 50\% and $\nu_0=\nu_-(\lambda_0)$ with the probability of 50\%. We numerically solve Eq. (\ref{eq:02}) and obtain the values of $\Omega_\phi$ and $w_\phi$ at $N=88.13$. Then we figure out which bin the obtained point $(\Omega_\phi,w_\phi)|_{N=88.13}$ belongs to.
  \item We repeat the previous step 11765 times, randomly plot 60 data points of $(\Omega_\phi,w_\phi)|_{N=88.13}$ in yellow and do a statistical analysis of the whole result. The points lie on the circle as we expected. The proportion is represented by the gray scale of each bin on the circle. In the inserted histogram, we plot the result versus the equivalent $N_{\rm eq}$. The vertical dashed line corresponds to $N_{\rm eq}=81.94$, at which $\Omega_\phi=0.7$ and $w_\phi\approx-1$ in Fig. \ref{fig:02} (b).
\end{enumerate}
This figure shows that the equivalent $N_{\rm eq}$ is distributed approximately uniformly in the whole period. The maximum value of the probability density is only about twice its minimum value. The worst result that most of the equivalent $N_{\rm eq}$ is distributed in a range far away from $N_{\rm eq}=81.94$ do not appear. In contrast, there are considerable points distributed around $N_{\rm eq}=81.94$. This result can be used to answer the question raised in Paper I: Whether the available initial condition and parameter settings are widespread? The answer is yes. No initial conditions and parameters need to be fine-tuned to explain the observed cosmic late-time acceleration.

\section{Properties in frequency domain}\label{sec:03}
\begin{figure*}[t]
  \centering
  \includegraphics[width=0.99\linewidth]{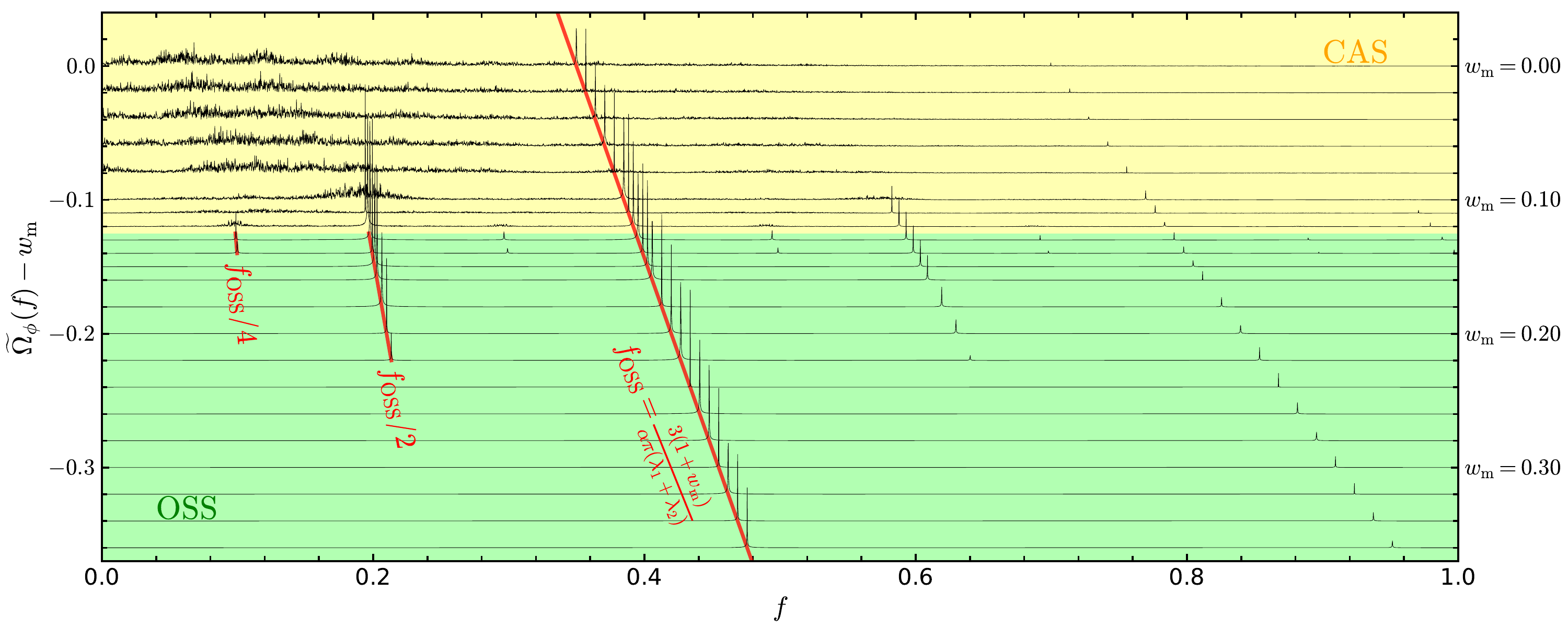}
  \caption{Fourier transform of $\Omega_\phi(N)$. The model parameters are $\lambda_1=4.5$, $\lambda_2=0.05$, $\alpha=0.6$ and $w_{\rm m}$ is denoted in the right $y$-axis. The initial conditions are the same as in Fig. \ref{fig:01}. We numerically solve Eq. (\ref{eq:02}) in $N\in[0,5000]$ and perform Fourier transform in $N\in[50,5000]$, which can effectively eliminate the dependence of the result on the initial conditions. We plot the principle frequency $f_{\rm OSS}$, the first period-doubling frequency $f_{\rm OSS}/2$ and the second period-doubling frequency $f_{\rm OSS}/4$ in red. The green region corresponds to OSS and the orange region corresponds to CAS.}
  \label{fig:05}
\end{figure*}

For the dynamical system Eq. (\ref{eq:02}), Fig. \ref{fig:01} shows the existence of OSS and CAS. In this section, we further study the system in frequency domain. Hereafter we regard $w_{\rm m}$ as a constant instead of Eq. (\ref{eq:04}). In Fig. \ref{fig:05}, we plot the $\widetilde{\Omega}_\phi(f)$, which is the Fourier transform of $\Omega_\phi(N)$, for $w_{\rm m}$ from $0.36$ to $0$. Other parameter and initial condition settings can be found in the caption. The difference between OSS and CAS becomes clear in frequency domain. OSS is periodic and its Fourier transform has only some equidistant peaks. CAS is chaotic and its Fourier transform is nonzero for a wide frequency range. For OSS, we find a principal frequency
\begin{equation}\label{eq:06}
  f_{\rm OSS}=\frac{3(1+w_{\rm m})}{\alpha\pi(\lambda_1+\lambda_2)},
\end{equation}
which exists for all parameter settings we consider [$\lambda_1+\lambda_2>4$, $0<\lambda_2<0.39$, $\alpha=\mathcal{O}(1)$ and $0\leqslant w_{\rm m}\leqslant1/3$]. This result is first obtained by numerical calculations, and then verified by analytical calculation in the limit of $\lambda_1-\lambda_2\ll1$ (see Appendix \ref{sec:A}). As $w_{\rm m}$ decreases from $0.36$ to $0.12$, the frequency $f_{\rm OSS}/2^n\,(n=1,2,3\cdots)$ appears in sequence. The CAS occurs when $w_{\rm m}\approx0.12$. This is exactly the route of period-doubling bifurcation to chaos \cite{Strogatz2018.book}. In Table \ref{tab:01}, we provide high-precision values of the critical $w_{\rm m}$ and see that the ratio $r_n$ converges to the Feigenbaum constant $4.669\cdots$ \cite{Feigenbaum1978.JSP.19.25}.
\begin{table}[!t]
  \centering
  \caption{The critical parameter $w_{\rm m}$ at which the period-doubling bifurcation occurs.  Other parameter settings are the same as in Fig. \ref{fig:05}. Here the ratio $r_n\equiv[w_{\rm m}(n)-w_{\rm m}(n-1)]/[w_{\rm m}(n+1)-w_{\rm m}(n)]$. In theory, we should have $\lim_{n\rightarrow\infty}r_n=4.669\cdots$, which is the Feigenbaum constant.}
  \label{tab:01}
  \begin{tabular*}{\hsize}{@{\ \ }@{\extracolsep{\fill}}lll@{\ \ }}
    \hline\hline
    $n$ & $w_{\rm m}$ & $r_n$  \\
    \hline
    1 & 0.2233 & --- \\
    2 & 0.1461 & 4.568 \\
    3 & 0.1292 & 4.605 \\
    4 & 0.12553 & 4.675 \\
    5 & 0.124745 & 4.673 \\
    6 & 0.124577 & \ldots \\
    \hline\hline
  \end{tabular*}
\end{table}

In Paper I, we pointed out that the critical points of the dynamical system Eq. (\ref{eq:02}) do not provide any quantitative information about its evolution. Here, we clarify the mathematical essence of the system: A route of period-doubling bifurcation to chaos. This is useful for a thorough understanding of this system in the future.

\section{Discussion}\label{sec:04}
In Paper I, we proposed a new dark energy model to approach the cosmological coincidence problem through multiacceleration scenario. Based on the calculations performed in the matter era, we pointed out that the model is able to explain the observed cosmic late-time acceleration. Meanwhile, we observed that the system with reasonable parameters exhibits as chaos, which brings technical difficulty to the global cosmological parameter constraints. In this paper, we revisit the cosmological background dynamics after considering the early radiation component. Our result suggests that the theoretically preferred Universe evolves as OSS in the radiation era and as CAS in the matter era. In this scenario, the above technical difficulty disappears and all the advantages discussed in Paper I are preserved. Observationally, we find that the model can well fit the Hubble parameter data in the range of $z\lesssim2$, and the upcoming 21\,cm signal can distinguish our model from the standard $\Lambda$CDM model. In addition, the mathematical essence of the dynamical system is identified as a route of period-doubling bifurcation to chaos.

So far, our discussion has mainly focused on the cosmological background evolution. In modern cosmology, the theoretical and observational aspects about cosmological perturbations are very important. Here, we briefly discuss the issues about cosmological scalar and tensor perturbations in the oscillating dark energy model. Cosmological scalar perturbation is related to the large scale structure formation and the anisotropy of the cosmic microwave background (CMB). For six dark energy models characterized by parameterized oscillating EoS, Ref. \cite{Pace2012.MNRAS.422.1186} pointed out that the structure formation is unable to distinguish these models from the concordance $\Lambda$CDM model. Refs. \cite{Pan2018.PRD.98.063510,Tamayo2019.MNRAS.487.729} analyzed the CMB anisotropy and performed the global parameter constraints for five parameterized oscillating dark energy models. Their results show that most of the parameterized oscillating dark energy models have insignificant effects on the CMB anisotropy. Therefore, the parameterized oscillating dark energy model can well fit the observations related to scalar perturbation. We expect our model can do the same and quantitative analysis will be given in the future works. For the tensor perturbation, we focus on the primordial gravitational waves (PGWs) and its influence on the $B$-mode polarization spectra of CMB. In the radiation era of the standard $\Lambda$CDM model, the PGW amplitude always decreases because the radiation-dominated Universe is decelerating. This property can be preserved in our model with suitable parameter settings (see $w_{\rm tot}$, which is related to $\ddot{a}/a$, in Fig. \ref{fig:01}; Note that multiacceleration scenario appears only in the matter era is also sufficient to solve the cosmological coincidence problem). This may be necessary to explain the small tensor-to-scalar ratio obtained through $B$-mode observations \cite{Akrami2018.arXiv.1807.06211}. Detailed analysis on this issue will be presented in the future.

\section*{Acknowledgements}
This work was supported by the Initiative Postdocs Supporting Program under Grant No. BX20200065.

\appendix
\section{Analytical verification of Eq. (\ref{eq:06})}\label{sec:A}
In Sec. \ref{sec:03}, we obtained Eq. (\ref{eq:06}) based on numerical results. Here we provide an analytical verification in the limit of $\lambda_1-\lambda_2\ll1$. The parameter settings for analytical and numerical calculations are different. However, our result shows that Eq. (\ref{eq:06}) is valid for a wide parameter space. For convenience, we use $\varepsilon$ to denote a mathematical infinitesimal, and we set $\varepsilon=1$ after the Taylor expansion. All the following calculations are only valid to $\mathcal{O}(\varepsilon)$ level. The OSS is periodic and can be written as a Fourier series. Therefore, for the variables, we may assume
\begin{subequations}\label{eq:A1}
\begin{align}
  x_1&=a_{10}+\varepsilon\left[a_{11}\cos(2\pi f_{\rm oss}N)+b_{11}\sin(2\pi f_{\rm oss}N)\right]\nonumber\\
  &\quad+\mathcal{O}(\varepsilon^2),\\
  x_2&=a_{20}+\varepsilon\left[a_{21}\cos(2\pi f_{\rm oss}N)+b_{21}\sin(2\pi f_{\rm oss}N)\right]\nonumber\\
  &\quad+\mathcal{O}(\varepsilon^2),\\
  \lambda&=a_{30}+\varepsilon\left[a_{31}\cos(2\pi f_{\rm oss}N)+b_{31}\sin(2\pi f_{\rm oss}N)\right]\nonumber\\
  &\quad+\mathcal{O}(\varepsilon^2),\\
  \nu&=a_{40}+\varepsilon\left[a_{41}\cos(2\pi f_{\rm oss}N)+b_{41}\sin(2\pi f_{\rm oss}N)\right]\nonumber\\
  &\quad+\mathcal{O}(\varepsilon^2).
\end{align}
\end{subequations}
The cosmological scaling solution \cite{Copeland1998.PRD.57.4686} gives the background values: $a_{10}=\sqrt{3/2}(1+w_{\rm m})/\lambda_{12}$, $a_{20}=\sqrt{3(1-w_{\rm m}^2)/(2\lambda_{12}^2)}$, $a_{30}=\lambda_{12}$, $a_{40}=0$, where $\lambda_{12}=(\lambda_1+\lambda_2)/2$. The constraint equation (\ref{eq:03}) can be written as
\begin{equation}
  (\lambda_1+\lambda_2-2\lambda)^2+\frac{2}{3}\alpha^2\nu^2=\varepsilon^2(\lambda_1-\lambda_2)^2.
\end{equation}
which is automatically satisfied for Eq. (\ref{eq:A1}) in the level of $\mathcal{O}(\varepsilon)$. Substituting Eq. (\ref{eq:A1}) into Eq. (\ref{eq:02c}), we obtain
\begin{subequations}\label{eq:A3}
\begin{align}
  a_{41}&=\frac{\sqrt{6}\pi}{3}\frac{b_{31}f_{\rm oss}(\lambda_1+\lambda_2)}{1+w_{\rm m}},\\
  b_{41}&=-\frac{\sqrt{6}\pi}{3}\frac{a_{31}f_{\rm oss}(\lambda_1+\lambda_2)}{1+w_{\rm m}}.
\end{align}
\end{subequations}
Substituting Eq. (\ref{eq:A1}) into Eq. (\ref{eq:02d}), we obtain
\begin{subequations}\label{eq:A4}
\begin{align}
  a_{41}&=\frac{3\sqrt{6}b_{31}(1+w_{\rm m})}{\alpha^2\pi f_{\rm oss}(\lambda_1+\lambda_2)},\\
  b_{41}&=-\frac{3\sqrt{6}a_{31}(1+w_{\rm m})}{\alpha^2\pi f_{\rm oss}(\lambda_1+\lambda_2)}.
\end{align}
\end{subequations}
The combination of Eqs. (\ref{eq:A3}) and (\ref{eq:A4}) gives Eq. (\ref{eq:06}).

%

\end{document}